\documentclass[aps,prb,twocolumn,showpacs]{revtex4}

\usepackage{mathptm}    
\usepackage{dcolumn}                    
\usepackage{bm}                         
\usepackage{graphicx}

\usepackage{times}

\newcommand{\PRG}{PuRhGa$_5$}
\newcommand{\PCG}{PuCoGa$_5$}

\begin{document}

\title{The nuclear spin-lattice relaxation rate in the PuMGa$_5$ materials}

\author{Yunkyu Bang}
\affiliation{Department of Physics, Chonnam National University,
Kwangju 500-757, and Asia Pacific Center for Theoretical Physics, Pohang 790-784, Korea}

\author{M. J. Graf, N. J. Curro, and A. V. Balatsky}
\affiliation{Los Alamos National Laboratory, Los Alamos, New
Mexico 87545, USA}

\begin{abstract}
We examine the nuclear spin-lattice relaxation rates $1/T_1$ of
\PRG\ and \PCG, both in the superconducting and normal
states, as well as their Knight shifts.
Results for both compounds are consistent with a superconducting gap of d-wave symmetry
in the presence of strong impurity scattering, though with quite different
gap-over-$T_c$ ratios $2 \Delta_{0}/ k_B T_c$ (8 for \PCG\ and 5 for \PRG).
In the normal state,  \PRG\ exhibits a gradual suppression of $(T_1
T)^{-1}$  below 25 K, while measurements for \PCG\ reveal a monotonic increase down
to $T_c$. 
We propose that this behavior is consistently understood by the crossover from the
two-dimensional quantum antiferromagnetic regime of the local $5f$-electron
spins of Pu to the concomitant formation
of the fermion pseudogap based on the two-component spin-fermion model.
\end{abstract}


\date{\today}
\maketitle

\section{Introduction}

The discovery of superconductivity (SC) in plutonium based systems such as
\PCG\ and \PRG\ has stimulated the study of unconventional
superconductivity and the pairing symmetry and mechanism in these materials.\cite{Sarrao,Wastin}
The symmetry of an unconventional superconductor is reduced compared to the symmetry
of its normal state, thus resulting in many novel properties of the quasiparticle excitation spectrum.
With a record high superconducting transition temperature $T_c$ of the order of 20 K among
the $4f$ and $5f$ electron based compounds, this class of materials provides hope for a unifying
pairing mechanism from heavy fermion superconductors to the high-T$_c$ cuprates.\cite{Curro,Pu-pairing}
It is believed that the superconducting action in Pu-115 [Pu{\it M}Ga$_5$ with {\it M}=Co and Rh]
derives itself from the unique character of the $5f$ electrons of plutonium.\cite{Pu-pairing}
Pu{\it M}Ga$_5$ is isostructural to the tetragonal Ce-115 series [Ce{\it M}In$_5$].
Very recently,
Curro and coworkers \cite{Curro} proposed, based on their measurements of the Knight shift
and spin-lattice relaxation rates, that the Pu-115 compounds are bridging the
superconducting and normal-state properties of the heavy-fermion Ce-115 and high-temperature
copper-oxide superconductors. Therefore providing a means for tuning the interaction
strength of antiferromagnetic spin fluctuations to intermediate values between both extreme
limits.\cite{Bauer}

In this paper, we examine the spin-lattice relaxation rates $T_1 ^{-1}$
of \PCG\ and \PRG\ in the superconducting and normal states.
Very recently, Sakai and coworkers\cite{Sakai}
have measured  $T_1^{-1}$ in \PRG\ in the superconducting state and
found evidence for lines of nodes in the gap function, just as in \PCG, but
with a much reduced superconducting transition temperature
$T_c \approx $ 8.5 K versus $T_c \approx 18.6 $ K.
Both materials differ primarily in their lattice parameters,
since Rh has a slightly larger atomic radius than Co. \cite{Bauer}
For example, the $ab$ plane and $c$ axis lattice constants are larger in \PRG\, but
the ratio $c/a$ is larger in \PCG, suggesting that \PRG\ is less two dimensional (2D).
Predicting the consequences of this elongation  on the superconducting instability is not
straightforward; however, we find that this  small change of lattice constants leads to
dramatic changes in the electronic and magnetic properties in this class of materials.

Here, we give a detailed theoretical description of the spin-lattice relaxation rate
and predict what should be observed if the Knight shift was measured on the same sample.
Our self-consistent treatment of impurity scattering in the superconducting state
goes beyond the two-fluid approach used by Sakai et al. \cite{Sakai}, which was used to
explain the large residual density of states in \PRG.
Further, our theoretical fits to the experimental data of \PRG\ lead us to three important
conclusions:
First, the measured pair-breaking effect of impurities in  \PRG\ reduces the transition
temperature of a hypothetically pure sample by only 0.5 K. Therefore, the lower
$T_c$ of  \PRG\ is an intrinsic property and is not due to impurities.
Second, the theoretical fits indicate that the ratio
$2 \Delta_0 / k_B T_c$ is markedly reduced in \PRG\ ($\sim 5$) versus \PCG\ ($\sim 8$).
This fact indicates that the mediating bosonic pairing glue is stronger in \PCG.
Assuming that Pu-115 compounds are spin-fluctuation
mediated superconductors,\cite{Pu-pairing} we conclude that
Rh substitution reduces the strength of the mediating spin fluctuations.
Indeed, this conclusion is supported by the observed behavior of $1/T_1$ in the normal state of both
compounds.
Third and most interestingly, when the experimental data are plotted as $T_1 T$ versus $T$,
we find  that in the normal state $T_1 T$ of \PRG\ saturates and is nearly flat over a wide
temperature region  $T_c < T < 3 T_c$,
whereas $T_1 T$  in \PCG\ shows a monotonic decrease down to $T_c$.
This saturation resembles closely the pseudogap (PG) feature of the underdoped cuprates.\cite{Magishi}
As a possible explanation of this phenomenon, we propose the two-component spin-fermion model of
antiferromagnetically correlated metals\cite{Bang 02} and argue that the two natural
energy gaps accounting for the PG behavior in Pu-115 are the spin gap $\Delta_{SG}$
and the fermion gap $\Delta_{PG}$.

\section{Superconducting State}

The experimental techniques of nuclear magnetic resonance (NMR) and nuclear quadrupolar
resonance (NQR) have been used successfully in the past to distinguish between the
spin states of Cooper pairs (spin singlet vs. spin triplet pairing) and provide indirect
information on the symmetry of the gap function -- fully gapped vs.  nodal lines or nodal
points in the gap function on the Fermi surface.
Both techniques probe directly the quasiparticle density of states and reveal indirect
information about the pairing symmetry.

The standard explanation of power vs. exponential laws in the low-temperature behavior
of thermodynamic and transport properties, for example, the spin-lattice relaxation rate
$1/T_1$, comes from the difference of nodal and fully gapped excitation spectra in the
superconducting state.
In clean nodal superconductors $1/T_1$ exhibits a nearly $T^3$ behavior
far below the superconducting transition temperature, $T \ll T_{c}$, while it is exponential for
gapped superconductors.
On the other side,  deviations from this behavior, like the $T$-linear temperature dependence
of $1/T_1$ at low temperatures, are explained by impurity effects in an
unconventional superconductor with lines of nodes on the Fermi surface.

In our calculations the effect of impurity scattering is included within the self-consistent $T$-matrix
approximation,\cite{Bang04} which is the standard formulation for pointlike
defects in a superconducting dilute alloy.\cite{Hirschfeld86,SchmittRink86,Monien87}
For the case of particle-hole symmetry of the quasiparticle excitation spectrum
the Nambu component $T_3$ of the $T$ matrix vanishes,
and for a d-wave order parameter with isotropic scattering $T_1=0$
(also without loss of generality we can choose $T_2=0$ by general U(1) gauge
symmetry), where $T_i$ is the $i$th component of the $2\times 2$
Nambu matrix expanded in Pauli matrices. Then we need to calculate only
$T_0(\omega)$. The impurity self-energy is given by
$\Sigma_{0}=\Gamma T_{0}$, where $\Gamma=n_i/\pi N_{0}$. Here $N_0$ is
the normal density of states (DOS) at the Fermi surface, $n_i$ is the impurity
concentration;  $T_0 (\omega_n) =\frac{g_0 (\omega_n)}{[c^2-g_0 ^2
(\omega_n)]}$, where $g_0 (\omega_n) = \frac{1}{\pi N_0}  \sum_k
\frac{i \tilde{\omega}_n}{\tilde{\omega}_n^2 + \epsilon_k^2
+\Delta^2(k)}$.
The impurity renormalized Matsubara frequency is defined by
 $\tilde{\omega}_n=\omega_n+\Sigma_0$, with $\omega_n=\pi T (2n+1)$,
and the scattering strength parameter $c$ is related to the s-wave phase
shift $\delta_0$  by $c=\cot(\delta_0)$.
Using this self-energy $\Sigma_0$ the following gap equation is solved self-consistently,

\begin{eqnarray}
\label{eq_gap}
\Delta(\phi) &=& - N_0 \int \frac{d \phi^{'}}{2 \pi}
V(\phi-\phi^{'})
\nonumber \\ &\times&
  T \sum_{\omega_n} \int^{\omega_c}_{-\omega_c}
  \frac{d \epsilon
    \Delta(\phi^{'})}{\tilde{\omega}_n^2 + \epsilon^2
    +\Delta^2(\phi^{'})}
\ ,
\end{eqnarray}
where $V(\phi-\phi')$ is the angular parameterization of the
pairing interaction, and $\omega_c$ is a typical cutoff energy.
We assume the canonical d-wave gap function of the form
$\Delta(\vec{k})= \Delta_0 (\cos k_x -\cos k_y)$ or
$\Delta(\phi) = \Delta_0 \cos(2 \phi)$ for a cylindrical Fermi surface.
The pairing potential $V(\phi-\phi^{'})$  induces a gap with d-wave symmetry.
Although its microscopic origin is not the issue of this paper, we
believe it originates from  2D antiferromagnetic (AFM) spin fluctuations.
The static limit of the spin susceptibility of the AFM fluctuations,
$\chi({\bf q},\omega=0) \sim \frac{1}{(\bm{q}-\bm{Q})^2+\xi^{-2}}$,
is parameterized near the AFM wave vector $\bm{Q}$ as \cite{Bang 02}
\begin{equation}
\label{eq_pairing}
V(\phi-\phi^{'})=V_d(b) \frac{b^2}{(\phi-\phi^{'} \pm \pi/2)^2+b^2}
\ ,
\end{equation}
where the parameter $b$ is inverse proportional to the AFM correlation
length $\xi$, normalized by the cylindrical Fermi surface $(\xi \sim a \pi/ b$; $a$ is the
lattice parameter). For all calculations in this paper, we choose
$b=0.5$ which is not a sensitive parameter for our results unless
$\xi$ is very large ($ b < 0.1$),\cite{Bang 02} i.e.,  within the
range of $0.1 < b < 1$ our results show little variations and
are qualitatively the same.

With the gap function $\Delta(\phi)$ and  $T_0 (\omega)$ obtained from Eq.~(\ref{eq_gap})
(where $T_0 (\omega)$ is analytically continued from $T_0 (\omega_n)$ by Pad\'e
approximation) we calculate the nuclear spin-lattice relaxation
rate  $1/T_1$ following the standard formulation\cite{Hirschfeld86,Bang04,Curro}
\begin{eqnarray}
\label{eq_spin_lattice_rate}
\frac{1}{T_1 T} &\sim&  -\int_0 ^{\infty} \frac{\partial f_{FD}
(\omega)}{\partial \omega}
\left[
 \left\langle Re
   \frac{\tilde{\omega}}{\sqrt{\tilde{\omega}^2-\Delta^2(\phi)}}
  \right\rangle_{\phi}^2
\right.
\nonumber \\ &&
\left.
 + \left\langle Re
     \frac{\Delta(\phi)}{\sqrt{\tilde{\omega}^2-\Delta^2(\phi)}}
    \right\rangle_{\phi}^2
\right],
\end{eqnarray}
and the superconducting spin susceptibility $\chi_S$
\begin{eqnarray}
\label{eq_spin_susceptibility}
\frac{\chi_S}{T} &\sim&  -\int_0 ^{\infty} \frac{\partial f_{FD}
(\omega)}{\partial \omega}
 \left\langle Re
   \frac{\tilde{\omega}}{\sqrt{\tilde{\omega}^2-\Delta^2(\phi)}}
  \right\rangle_{\phi}
,
\end{eqnarray}
where $f_{FD}(\omega)$ is the Fermi-Dirac function, the impurity renormalized
quasiparticle energy
$\tilde{\omega}=\omega+\Sigma_0(\omega)$, and
$\langle...\rangle_{\phi}$ means the angular average over the Fermi surface. The first
term in the bracket of Eq.~(\ref{eq_spin_lattice_rate})
is  $N^2 (\omega)$. The second term vanishes in
our calculations because of the symmetry of the gap function.  To calculate $1/T_1 T$
using Eq.~(\ref{eq_spin_lattice_rate}), or $\chi_S$ using Eq.~(\ref{eq_spin_susceptibility}),
we need the full temperature dependent gap function
$\Delta(\phi,T)$ and $T_{c}$. Our gap equation Eq.~(\ref{eq_gap}) is the BCS
gap equation, therefore it gives  the BCS temperature behavior for
$\Delta(\phi,T)$ and $\Delta_0=2.14 \, {k_B T_c}$ for the standard
weak-coupling d-wave SC. In order to account for strong-coupling effects
we use the phenomenological formula
$\Delta(\phi,T)=\Delta(\phi,T=0)~ \Xi(T)$ with $\Xi(T)=\tanh (\beta
\sqrt{T_{c}/T-1})$, and parameters $\beta$ and $\Delta_0/ T_{c}$.   Then we
only need to calculate $\Delta(\phi,0)$ at zero temperature. The temperature
dependence of $\Sigma_0(\omega,T) \equiv \Gamma T_0 (\omega,T)$ is similarly
extrapolated: $T_0(\omega,T)=T_0(\omega,T=0)~ \Xi(T) + T_{normal}(1-\Xi(T))$,
where $T_{normal}=\Gamma/(c^2+1)$ is the normal state $T_0$. In our numerical
calculations we chose $\beta=1.74$, because our final results are not very
sensitive with respect to this parameter,
while the ratio $\Delta_0/ k_B T_{c}$ is an important parameter
to simulate strong-coupling effects.  The larger the gap ratio the more
important are strong-coupling effects.

\begin{figure}
\noindent
\includegraphics[width=100mm]{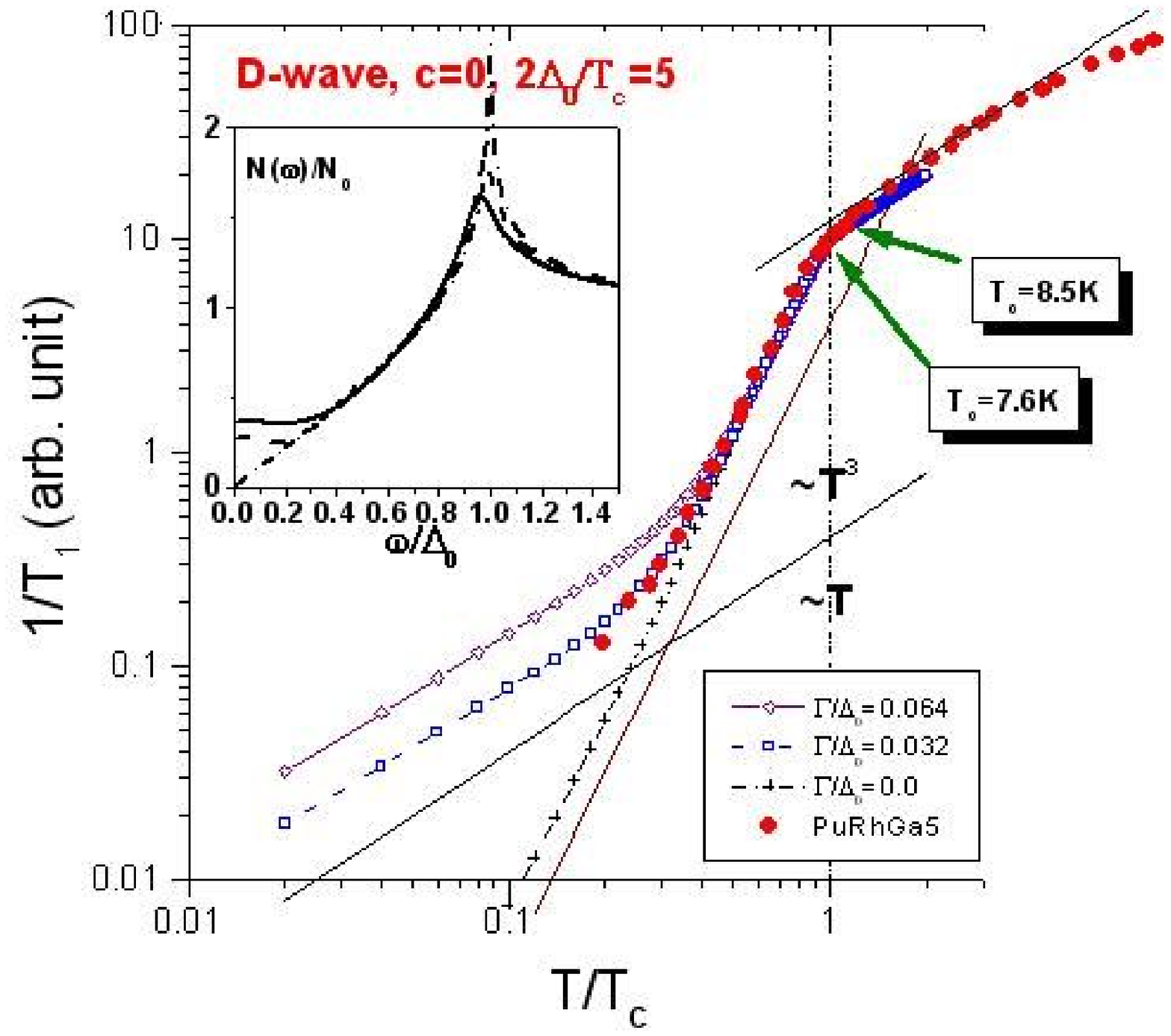}
\caption{The NQR spin-lattice relaxation rate \cite{Sakai} plotted versus
temperature normalized by the superconducting transition
temperature $T_c=7.6$ K. Calculations are shown for $2\Delta_0 = 5 \,
k_B T_c$ and three values of the impurity scattering rate $\Gamma$
for unitary scattering. Inset: The normalized quasiparticle
density of states is shown for corresponding values of the
impurity scattering rate $\Gamma/\Delta_0 = 0, 0.032, 0.064$.}
\label{fig1}
\end{figure}

\begin{figure}
\noindent
\includegraphics[width=100mm]{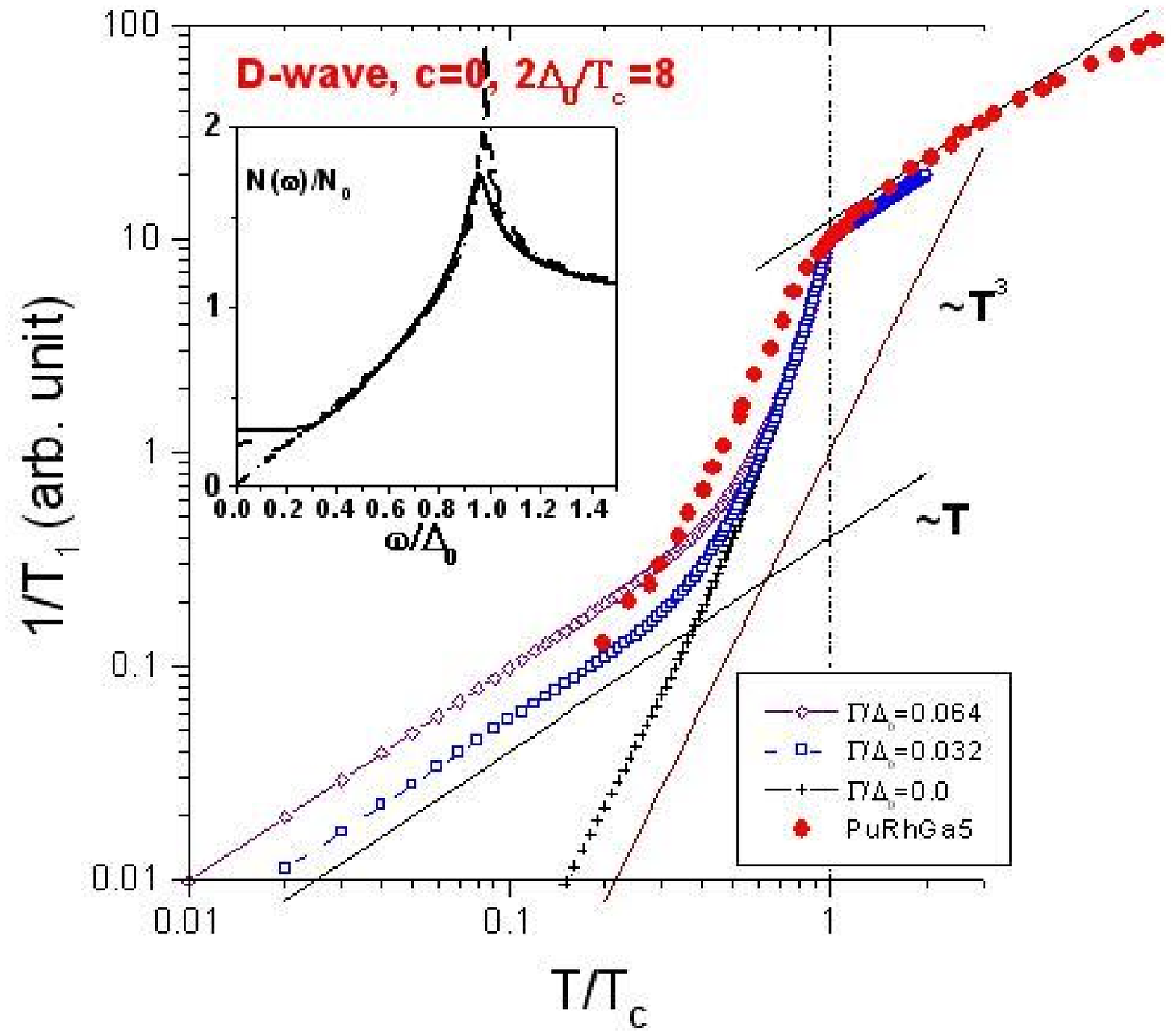}
\caption{The NQR spin-lattice relaxation rate \cite{Sakai} plotted versus
temperature normalized by the superconducting transition
temperature $T_c=7.6$ K. Calculations are shown for $2\Delta_0 = 8 \,
k_B T_c$ and three values of the impurity scattering rate $\Gamma$
for unitary scattering. Inset: The normalized quasiparticle
density of states is shown for corresponding values of the
impurity scattering rate $\Gamma/\Delta_0 = 0, 0.032, 0.064$.}
\label{fig2}
\end{figure}

\begin{figure}
\noindent
\includegraphics[width=100mm]{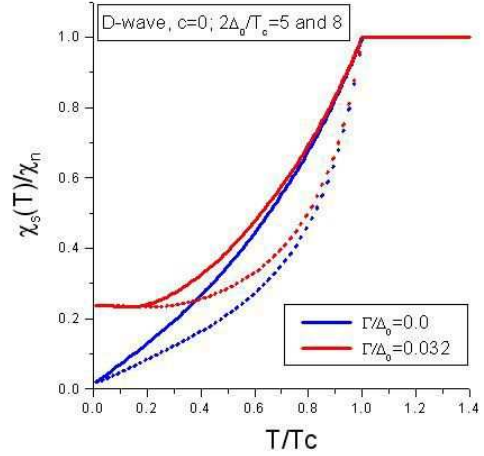}
\caption{The calculated spin susceptibility $\chi_S$ of a d-wave
SC normalized by its normal state value $\chi_{N}$ for gap values
$2\Delta_0 = 5 \,{k_B T_c}$ (solid lines) and $8 \,{k_B T_c}$
(dotted lines), and impurity scattering rates $\Gamma/\Delta_0 =
0$ and $0.032$. } \label{fig3}
\end{figure}

In figures \ref{fig1} and \ref{fig2} the spin-lattice relaxation rate of
\PRG\ by Sakai et al.~\cite{Sakai} is shown. For ease of comparison $1/T_1$ was normalized
to $1/T_1=10$ at $T=T_{c}$.
The insets show the corresponding normalized quasiparticle DOS for varying scattering rates $\Gamma$.
With our earlier described choice of gap parameters ($\beta=1.74$ and $2\Delta_0=5 \,{k_B T_c}$)
an impurity scattering rate of $\Gamma/\Delta_0=0.032$ in the unitary limit ($c=0$) is
enough to completely fill the low energy gap with impurity states,
where $N(\omega=0)$ reaches more than 25\% of the normal-state DOS $N_0$.
For a higher value of $\Gamma/\Delta_0=0.064$, the $T$-linear region extends up to $\sim 0.35$
$T_{c}$. At temperatures near $T_{c}$ the coherence peak is almost invisible
because of the sign-changing gap function, i.e., vanishing of the second term in
Eq.~(\ref{eq_spin_lattice_rate}).

Below $T_{c}$ the spin-lattice relaxation
rate shows a nearly $T^3$ behavior until it crosses over to a $T$-linear region.
A comparison with the experimental data by Sakai et al.\cite{Sakai} is in good agreement
with a scattering rate close to $\Gamma/\Delta_0=0.032$.
Based on this value, we estimate the reduction of $T_{c0}$ of a hypothetically pure sample
due to impurities to be $T_{c0} - T_c = \frac{\pi}{4} \Gamma \approx 0.53$ K,
which results in $T_{c0}\approx 8.1$ K or $9.0$ K, depending on the value of $T_c = 7.6$ K or $8.5$ K.
This small suppression and large impurity induced DOS is a characteristic of unitary impurities (c=0).

In Fig.~\ref{fig1} we obtained a better fit to the experimental data (symbols)
assuming a slightly lower superconducting transition temperature
$T_c = 7.6$ K than the reported value of $T_c = 8.5$ K by Sakai et al.\cite{Sakai}
This could indicate the presence of a
pseudogap similar to the high-temperature superconductor YBa$_2$Cu$_3$O$_{7-\delta}$,
where $1/T_1$ is suppressed starting above $T_c$.

In Fig.~\ref{fig2}, we plot $1/T_1$ for an enhanced strong-coupling
d-wave gap of $2\Delta_0=8\,{k_B T_c}$, as was recently reported for \PCG.\cite{Curro}
Due to the larger gap value, the theoretical $1/T_1$ drops initially faster below $T_c$ than
the experimental data.  Hence, we find a poorer fit to the measured data for this choice of
strong-coupling gap.  It demonstrates that the comparison of $1/T_1$ data with
theoretical calculations is a useful tool for determining
the strong-coupling gap value $\Delta_0/k_B T_c$.

Fig.~\ref{fig3} shows the prediction for the spin susceptibility, $\chi_S$,  or its
corresponding NMR Knight shift, $K = K_0 + A \chi_S$, where $K_0$ and $A$ are constants for
most materials.
$\chi_S$ is calculated for the same d-wave gap values as was used for the spin-lattice
relaxation rates in figures \ref{fig1} and \ref{fig2}. Again a modest impurity scattering
rate of $\Gamma/\Delta_0 = 0.032$ results in a large residual susceptibility at zero temperature,
equivalent to roughly $25 \%$ of the normal-state DOS or spin susceptibility $\chi_N$.
The quantitative difference in the spin susceptibilities between gap values
$2\Delta_0 = 5 \, {k_B T_c}$ and $8\,{k_B T_c}$ should be easily discernible in
future measurements of the Knight shift.

At first sight it might appear that the sample \PRG\ had three times more defects
than \PCG\ of similar age,
based on our best-fit scattering rates of $\Gamma/\Delta_0=0.032$ for \PRG\ and
$\Gamma/\Delta_0=0.01$ for \PCG. This could be explained by slightly different
isotope mixes of plutonium used. However, if we express the scattering rates in
absolute values, we find very similar impurity scattering rates for both samples,
namely, $\Gamma = 0.6$ K for \PRG\ and $\Gamma= 0.7$ K for \PCG, consistent with
a common origin of defect generation due to self-irradiation by plutonium.

One final remark is warranted, namely, that the experimental data by Sakai et al.
in Fig.~\ref{fig1} are normalized assuming a slightly
lower superconducting transition temperature $T_c = 7.6$ K than
the reported value by the authors. Indeed, if $T$ was normalized by $T_c = 8.5$ K,
then the fitting would be poorer and in particular the excess relaxation rate
just below $T_c$ could not be explained by a simple superconducting transition.
The origin of this ambiguity of $T_c$ is not clear at the moment. If the true
$T_c$ of the sample was indeed 7.6 K, as was used in Fig.~\ref{fig1}, then the
incorrectly assigned $T_c=8.5$ K might be due to the presence of a
pseudogap that will be discussed in the next section.

\section{Normal State}

\begin{figure}
\noindent
\includegraphics[width=100mm]{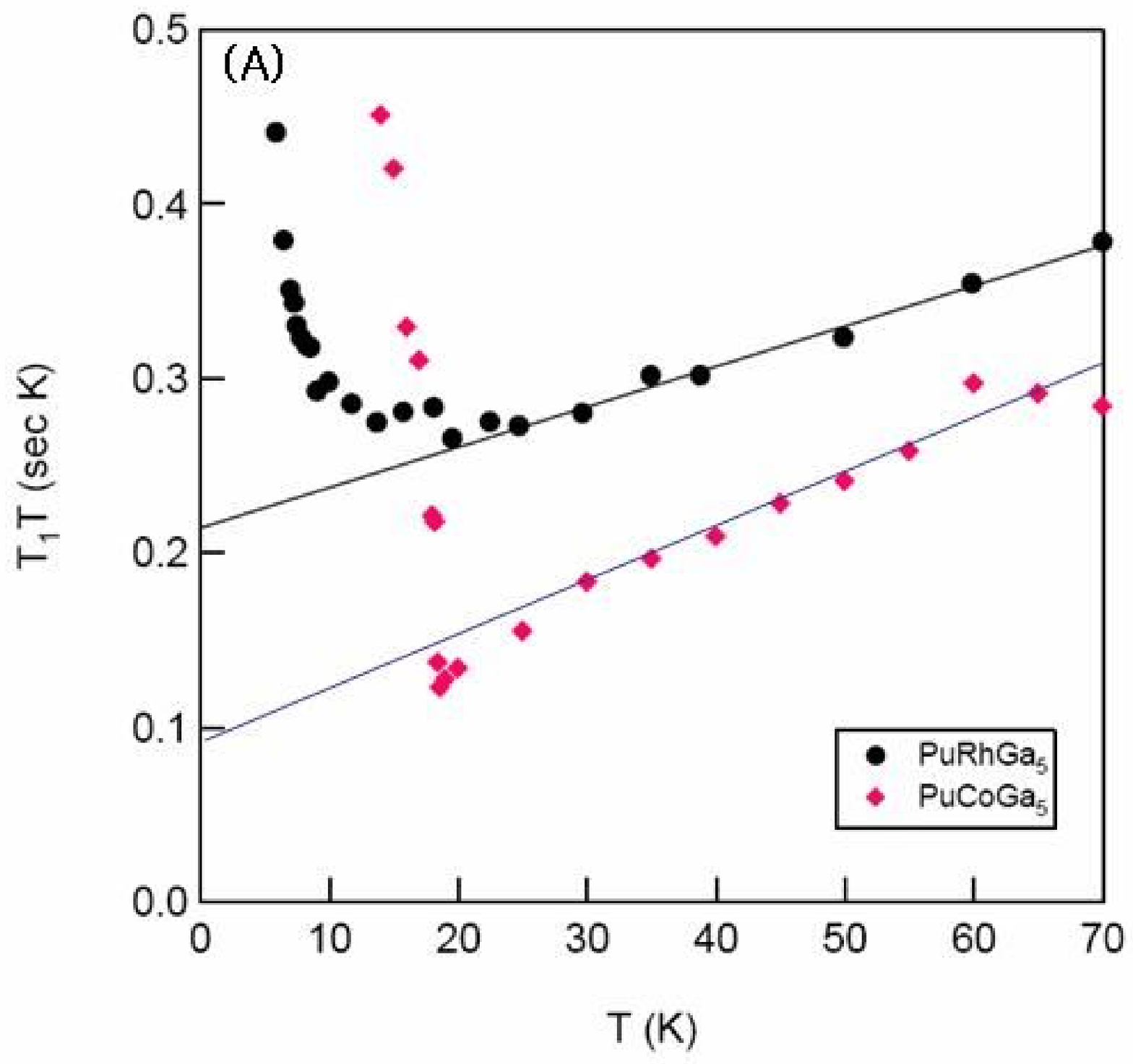}
\includegraphics[width=100mm]{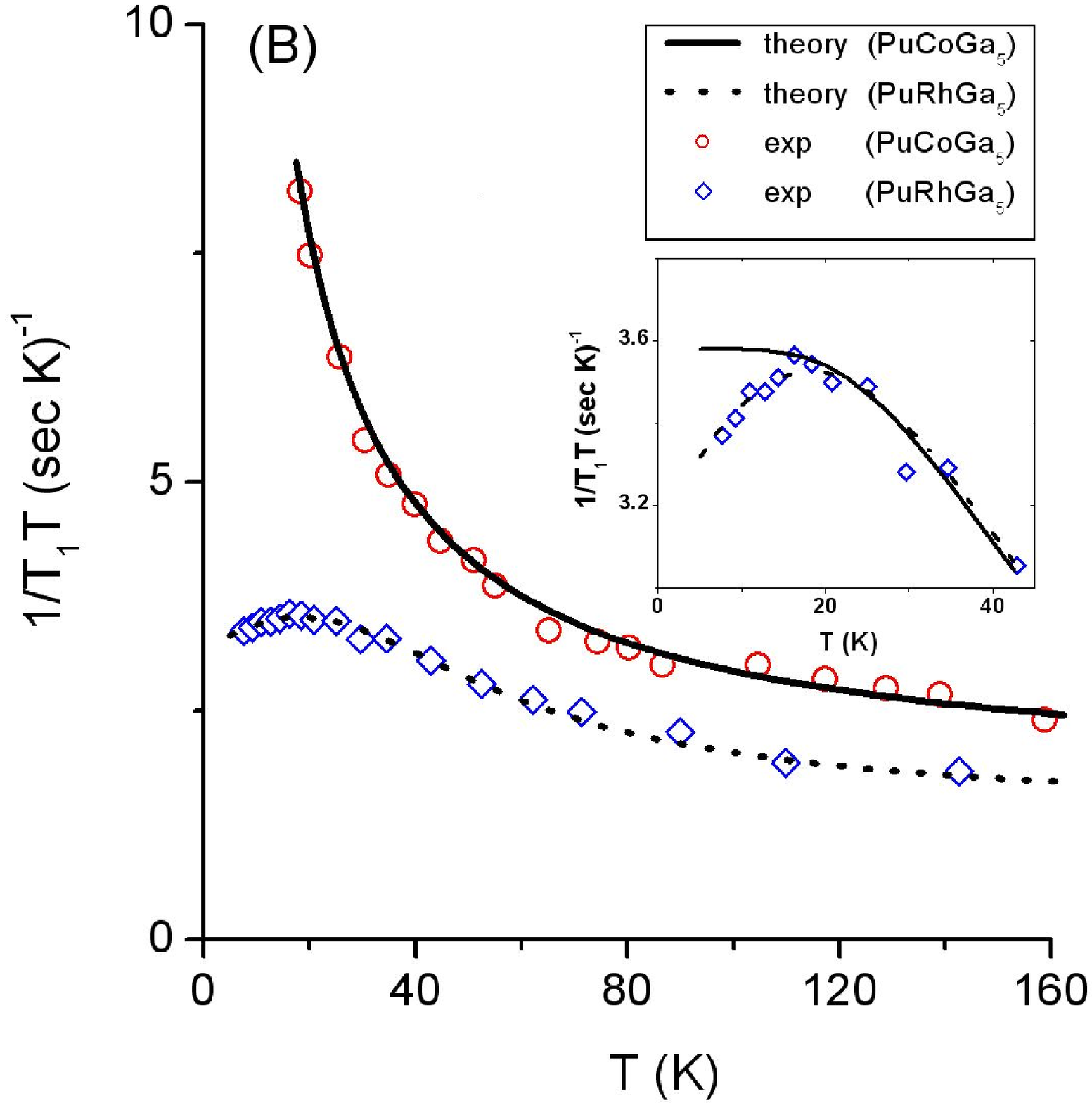}
\caption{(a) Plot of $T_1 T$ data vs. $T$ of \PCG\ and \PRG.
(b) Plot of $1/T_1 T$ vs. $T$ for a wider temperature range of the normal state for \PRG\ (open diamonds)
and \PCG\ (open circles) and the theoretical fits.  Parameters for the theoretical fits are
$\Delta_{SG}$ =20 K and $\Delta_{FPG}$ =8 K for \PRG\
and $\Delta_{SG}$ =0 K and $\Delta_{FPG}$ =8 K, respectively.
Inset: Close-up view for \PRG\ at low temperatures. Theoretical fits with
(dotted line) and  without (solid line) the fermionic PG correction
are shown for comparison. (\PRG\ data are from Ref.[6].)
}
\label{fig4}
\end{figure}

Next we address the non-Fermi liquid behavior of the Pu-115 compounds in the normal state.
As we will argue below, this behavior can be consistently understood within
spin fluctuation theory.
In Fig.~\ref{fig4}(a),  we plot the inverse of the measured $1/T_1$ for \PCG\ and \PRG\
multiplied by $T$ versus temperature up to higher temperatures (multiple times $T_c$).
A Fermi liquid should exhibit a constant $T_1 T$ in the normal state, in contrast to the
succinct features of Pu-115:
(1) at high temperatures ($T \gtrsim$ 25 K) both data
show $T$-linear behavior;
(2) $T_1 T$ in \PCG\  shows monotonic decrease  down to $T_c$ ---
a small deviation from the $T$-linear behavior in the region of
$T_c \lesssim  T  \lesssim$ 30 K calls
for additional explanation;
(3) the most interesting feature of the
curve is the gradual round-off of $T_1 T$ in the \PRG\ data below $\sim$
20 K. Even the superconducting transition at $T_c =$ 8.5 K or 7.6 K
is not clearly discernible from these data.
This roundoff in $1/T_1 T$ starting far above $T_c$ has been frequently observed in underdoped high
temperature superconductors and has been attributed to the suppression of
low-energy spin fluctuations associated with the pseudogap behavior.
Recently, experiments of several heavy fermion compounds showed such a pseudogap behavior. \cite{Sidorov,Mito}
While the pseudogap behavior in heavy fermions typically shows only in a very narrow temperature range of a few Kelvin, it
provides a much clearer evidence for its magnetic origin than in the cuprates.
For the cases of CeIn$_3$ and CeRhIn$_5$ \cite{Mito,Kawasaki} 
the pseudogap occurs in NMR data above the Ne\'el temperature $T_N$.
In addition, the pseudogap of CeCoIn$_5$ occurs above the superconducting transition, a system which is
known to be very close to the two-dimensional antiferromagnetic quantum criticality (QC).\cite{Sidorov}

Considering that Pu-115 materials are near a 2D antiferromagnetic
(AFM) instability, we start with the phenomenological model of the
antiferromagnetically correlated metal. The minimal set of low
energy degrees of freedom  are the fermionic charge excitations and
the collective spin excitations. This phenomenological theory is
also called spin-fermion model and has been intensively studied by
Pines and coworkers.\cite{Pines1}
In contrast to this standard spin-fermion model, we
proposed for heavy fermions the two-component spin-fermion model,\cite{Bang 02} 
where the spin modes originate directly from localized spins rather than
from collective particle-hole excitations. In a mixed momentum and
real-space representation the corresponding Hamiltonian is written
as
 \begin{equation}
 H = \sum_{{\bf k}, \alpha} c^{\dag}_\alpha({\bf
 k})\varepsilon({\bf k})c_\alpha({\bf
 k}) + \sum_{{\bf r},\alpha, \beta} J{\bf \vec{S}}({\bf r}) \cdot
c^{\dag}_\alpha({\bf r}){\bf \vec{\sigma}}_{\alpha \beta}c_\beta({\bf
 r}) + H_S
 \end{equation}
where the first term is the fermionic kinetic energy
and  the second term describes the coupling between local spins  ${\bf \vec{S}}({\bf r})$
and the spin density of the conduction electrons.
The last term  represents an effective low-energy Hamiltonian for the local spins.
When the local spins have a short range AFM correlation, the spin
correlation function has the general form \cite{Sachdev 95}
\begin{equation}
\chi ({\bf q},\omega) =
\frac{\chi({\bf Q}, 0)}{1 +  \xi^{2} |{\bf q} -{\bf Q}|^2
- \omega^2 / \Delta_{SG} ^2 - i\omega/ \bar{\omega}},
\end{equation}
\noindent
where  $\Delta_{SG}$ is the spin gap, ${\bf Q}$ the 2D AFM ordering vector,
$\xi$ the magnetic correlation length, and $\bar{\omega}$ the spin relaxation energy scale,
which comes from Landau damping of the fermionic sector.
Given the above form of the spin susceptibility $\chi({\bf q},\omega)$
and assuming the 2D AFM correlation \cite{2D AFM}, it has been shown that
1/T$_1$T.$\sim 1/\bar{\omega}$.  \cite{Sokol}
Further, it is known that $\bar{\omega} \sim \xi^{-1}$
for the z=1 quantum-critical phase of the 2D quantum antiferromagnet.\cite{CHN}
As was shown by Chakravarty and coworkers,\cite{CHN}
the magnetic correlation length displays, depending on the phases of quantum criticality (QC) or
quantum disorder (QD) the following behavior:
\begin{equation}
\xi^{-1} \sim \left\{
\begin{array}{ll}
$T$ & \textrm{for QC ~~($T > T^*$)} \\
\mbox{const.} & \textrm{for QD ~~($T < T^*$)}
\end{array}
\right.
\end{equation}

The high-temperature $T$-linear behavior in $T_1 T$, see Fig.~\ref{fig4}(a), of \PCG\ and \PRG\
can be understood as a generic feature of the quantum critical regime of the 2D Heisenberg antiferromagnet.\cite{CHN}
In principle, this high temperature QC explanation can be tested by the measurement of the spin-spin relaxation time
$T_{2G}$, as was done for the high-temperature superconductors;\cite{Curro97} then the simple relation
$T_1 T/T^z_{2G}$ =const.\  should characterize the QC dynamics in Pu-115 by determining the dynamic critical
exponent z.
Also the low-temperature round-off in $T_1 T$ of \PRG\ can be associated with the QC to QD crossover of the
2D AFM.
However, it is still not sufficient to explain the additional fall-off of $1/T_1 T$ between 20 K 
and $T_c\approx 7.6$ K (for more detail see the inset of Fig.~\ref{fig4}(b)).
We propose that this additional suppression of the spin-fluctuations is caused
by the suppression of the fermionic DOS (namely the fermionic pseudogap).
Since the term  $i \omega / \bar{\omega}$ in Eq.(6) originates from Landau damping of the fermionic sector,
instead one needs to include $\sim  i \omega N(E_{F},T) / \bar{\omega}$.
Therefore, when the fermionic DOS $N(E_F)$ becomes temperature dependent,
it needs to be included and leads to the modification $1/T_1 T \sim N(E_{F},T) / \bar{\omega}(T)$.

Several authors \cite{Fermion gap} have studied the influence of magnetic correlations on
fermionic quasiparticles  and found that increasing the magnetic correlation length $\xi$
causes a precursor effect of a spin-density wave state, which forms a quasi-gap in the DOS.
The amount of the suppression of the DOS depends sensitively on the parameters, such as the size of
the correlation length $\xi$, the coupling constant, and  temperature, etc.
In this paper, therefore, we merely introduce a phenomenological form of the pseudo-gapped DOS,
$N(E_{F},T,\Delta_{FPG})$. The systematic numerical studies of the fermionic pseudogap (FPG) $\Delta_{FPG}$ 
due to magnetic correlation
in the two-component spin-fermion model will be reported elsewhere.

Combining
the temperature dependence of the FPG and the magnetic
correlation of the 2D AFM, we can write $1/T_1 T$ in the
two-component spin-fermion model as
\begin{eqnarray}
1/T_1 T &=& A N_0 (E_{F})[1-\tanh^2(\frac{\Delta_{FPG}}{2 \sqrt{T^2+\Gamma^2}})] \\ \nonumber
& & \times \frac{1}{[\Delta_{SG} + T \exp (-4 \Delta_{SG} /T)]} +B  .
\end{eqnarray}

The first factor $N_0 (E_{F})[1 - \dots]$ is the phenomenological form of
the fermionic DOS with FPG $\Delta_{FPG}$ and damping rate $\Gamma$.
The second factor $\frac{1}{[\Delta_{SG} + \dots]}$ is a smooth crossover function \cite{CHN} for $\xi(T)$
describing the QC to QD behavior of Eq.(7) with the spin gap $\Delta_{SG} \sim T^*$.
Here, $B$ is a constant describing a temperature independent contribution and $A$ is a constant scale factor.
Using this formula, we fit the experimental data of \PCG\ and \PRG\ for normal state in Fig.~\ref{fig4}(b).
The fitted results are in excellent agreement with experiment, where the two key fitting parameters provide
estimates for the important energy scales of this phenomenological model.
For \PRG, we used $\Delta_{SG} = 20$ K and $\Delta_{FPG}=8$ K; the damping rate $\Gamma$ is not a very sensitive
model parameter, so we use $\Gamma$=25 K in all cases.  A spin gap of
$\Delta_{SG} \sim  T^* = 20$ K can be  read off from the data in 
Fig.~\ref{fig4}(a), where $T_1 T$ deviates from
a linear temperature dependence.
However, notice that without the FPG correction $\Delta_{FPG}$
the additional drop of $1/T_1 T$ in the region of  $T_c  < T < T^*$
cannot be explained only by the QC to QD crossover
of $\xi$, see the inset of Fig.~\ref{fig4}(b).
For \PCG, the monotonically increasing $1/T_1 T$ at lower temperatures implies increasing $\xi(T)$ and stronger
magnetic correlations than in \PRG.
Along the same line of thought, as  applied to \PRG, the FPG should also be formed in the case of \PCG.
In Fig.~\ref{fig4}(b),  we used $\Delta_{SG} = 0$ K and $\Delta_{FPG}=8$ K for \PCG.
The FPG effect is not much visible  because it is overwhelmed by the stronger temperature dependence of
$1/\bar{\omega} \sim \frac{1}{T}$ for \PCG.
For \PCG, we could obtain a same quality of good fits with different values of $\Delta_{FPG}$
as large as $\sim$ 20 K.

\begin{figure}
\noindent
\includegraphics[width=100mm]{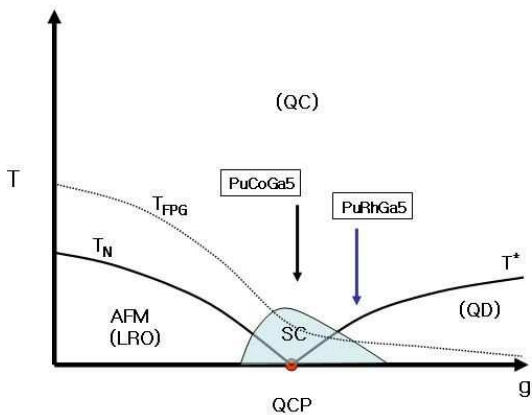}
\caption{The schematic phase diagram for the alloy system Pu(Co,Rh)Ga$_5$.
Thick solid lines are the 2D AFM transition temperature ($T_N$) and the
QC to QD crossover temperature ($T^*$).
The thin dotted line is the crossover temperature ($T_{FPG}$) to the fermionic
pseudogap (FPG) region.} \label{fig5}
\end{figure}

We summarize the effects of the magnetic correlations in \PCG\ and
\PRG\ on the thermodynamic behavior in the schematic phase diagram
in Fig.~\ref{fig5} with respect to a generic coupling parameter $g$.
The assignment of \PCG\ rather than \PRG\ to the vicinity of the
quantum critical point (QCP) in the QC region is consistent
with various other experiments: for instance, the lower $T_c$, the
smaller the value of the specific heat jump $\Delta C/T_c \approx
45$ mJ mol$^{-1}$ K$^{-2}$,\cite{Javorsky} and the smaller the
gap-over-$T_c$ ratio $2 \Delta_{0}/ T_c$ in \PRG\ compared to \PCG.
Also the resistivity of \PCG\ shows an anomalous power law
dependence \cite{Sarrao}, $\varrho \propto T^{4/3}$ , as expected near a quantum
phase transition. All these experimental data indicate that the 2D AFM
fluctuations are weaker in \PRG\ than in \PCG. In Fig.~\ref{fig5} we
added the crossover line (thin dotted line) to the FPG region, $T_{FPG}$. 
The formation of the FPG is not a
universal property and depends on material specific details.
Nevertheless, in order to understand the detailed behaviors of
$1/T_1 T$ in Pu{\it M}Ga$_5$ materials, it is indispensable to be
included.

To complete our discussion, we briefly describe possibility of alternative 
explanations.
In recent Knight shift (K) measurements in PuRhGa5, Sakai et al. \cite{private}, 
found that the Knight shift failed to track the bulk susceptibility ($\chi$)
below a temperature ~ 30K.  Such K-$\chi$ anomaly is known to exist in other heavy
fermion compounds and various explanations were proposed \cite{curro04}. Naturally this
anomaly is likely to be related to the anomalous temperature dependence of T$_1$ \cite{curro_prl}; however, 
the deviation of K $\propto \chi$ relation {\it per se} doesn't necessarily mean the suppression
of $1/T_1 T$.
There are three proposals for K-$\chi$ anomaly in the literature:
(1) CEF (crystal effective fields)\cite{curro01}: the population of different CEF levels of f-electron
changes with temperature and hence the HF (hyperfine) coupling between nuclei spin and the f-electron
spin obtains a temperature dependence.
(2) Kondo cloud screening \cite{Cox}: the HF coupling between nuclei spin and the f-electron
spin is mediated by the RKKY modulation of the conduction electrons. The onset of Kondo screening would
change the characteristics of the RKKY modulation and leads to the change of HF coupling.
(3) Two fluids model \cite{curro04}: total susceptibility consists of two part -- one from f-electrons and the
other from conduction electrons -- and each component has different temperature dependence
and  HF couplings to the nuclei spin.
Due to the different temperature dependence of each susceptibility, 
the non-proportionality between K and $\chi$ can be explained.

The proposals (1) and (2) are basically invoking on the temperature dependent HF coupling but by
different mechanisms: Kondo and CEF, respectively. 
Our model is similar to the proposal (3) in the spirit of the two fluids 
model. However, there are technical and conceptual differences between (3) and our model.
First, technical difference is the following.
While we also assume two susceptibilities -- one from local f-electron and the other from conduction electron,
our two fluids model is not a simple addition of two susceptibilities.
We assume that the dominant contribution ( the interesting temperature 
dependent part) always comes from the local f-electrons.
But the f-electron susceptibility essentially obtains its low energy dynamics through
coupling with the conduction electrons and as a result the conduction electron susceptibility
always feeds back into the f-electron susceptibility and vice versa (see $i \omega / \bar{\omega} $ 
term in Eq.(6)).
Second, the most important conceptual difference is that we assume that the temperature
dependence of the f-electron susceptibility for the range of interesting temperatures 
is arising from a magnetic correlation of local f-electrons but not 
from Kondo or CEF effect. 
This point of departure is very crucial and should be determined by experiments. 
There are already abundant experimental evidences 
that Ce{\it M} ({\it M} =Co, Rh, Ir)In$_5$ materials are inside or in the vicinity 
of  AFM ordered phase \cite{multi-phase}.
For Pu{\it M} ({\it M} =Co, Rh)Ga$_5$, there are yet no direct measurements of the magnetic correlations, 
but the fact of a D-wave pairing \cite{Curro} in the superconducting state and 
the anomalous temperature dependence of dc-resistivity \cite{Pu-pairing} indicate that PuMGa$_5$ materials
are near AFM instability as sketched in Fig.5.

\section{Conclusion}

We have studied the nuclear spin-lattice relaxation rates $1/T_1$ of
\PCG\ and \PRG, both in the normal and superconducting states.
In the superconducting state, both compounds display the features of
a dirty d-wave superconductor with impurity scattering in the unitary limit.
This also is borne out in the calculated Knight shifts.
The superconducting gap values of $2 \Delta_{0}$/ T$_c$ are $\sim 8$ and $\sim 5$ for
\PCG\ and \PRG, respectively, indicating that pairing fluctuations are much stronger
in \PCG\ than in  \PRG.

In the normal state, the temperature behavior of $1/T_1$ between both compounds is
qualitatively different at low temperatures.
While $1/T_1 T$ of \PCG\ displays a genuine quantum-critical behavior similar to
a 2D quantum antiferromagnet down to $T_c$,
\PRG\ shows a pseudogap-like suppression over a wide temperature
region from $T_c$ (7.6 K) to roughly $3 T_c $ (25 K).
Because of this remarkable observation, we proposed the two-component spin-fermion model
and argued that the magnetic spin gap originates on the local spins of the $5f$ electrons
of Pu and that the concomitant formation of the FPG ($\Delta_{FPG}$) can provide a consistent
explanation of these phenomena.

We argued previously\cite{Curro,Pu-pairing} that \PCG\ bridges the heavy fermion
superconductors and high-$T_c$ cuprates in terms of the superconducting pairing mechanism.
The observation of the pseudogap phenomenon in \PRG\ and its temperature
range of over two times $T_c$ (or roughly 15 K) is another evidence that Pu{\it M}Ga$_5$ is
indeed the missing link between heavy fermion superconductors and cuprates, holding the key ingredient
-- magnetic correlations -- yet with an intermediate energy scale.
To test and confirm this hypothesis, we propose systematic studies of the alloy system Pu(Co,Rh)Ga$_5$.
It will be an ideal system for further studies because it can be cleanly tuned to explore the magnetic
phase diagram sketched in Fig.~\ref{fig5}
without changing the carrier density in contrast to the high-$T_c$ cuprates.

\section{Acknowledgements}

We thank John Sarrao,  Joe Thompson, David Pines and Eric Bauer for many stimulating discussions.
Y. B. was supported by the KOSEF through the CSCMR and the Grant No. KRF-2005-070-C00044.
This research was supported by the U.S. Department of Energy at Los Alamos National Laboratory
under contract No.~W-7405-ENG-36.


\begin{references}
\bibitem{2D AFM}
We assume that the AFM correlation in PuMGa$_5$ is 2D like because
the band calculations and the crystal structure of  PuMGa$_5$ compounds,
isostructural to CeMCo$_5$ compounds,
indicate the 2D character of the electronic properties.
\end{references}
\end{document}